\documentclass[PHM, 2020]{PHMSociety}

\usepackage{amsmath,amssymb,amsfonts}
\usepackage{algorithmic}
\usepackage{graphicx}
\usepackage{textcomp}
\usepackage{xcolor}
\def\BibTeX{{\rm B\kern-.05em{\sc i\kern-.025em b}\kern-.08em
    T\kern-.1667em\lower.7ex\hbox{E}\kern-.125emX}}

\usepackage{amsmath,bm}

\newtheorem{definition}{Definition}

\usepackage[ruled,vlined]{algorithm2e} 
\usepackage{multirow}
\usepackage{mathtools}

\begin{document}

\title{Data-driven Residual Generation for Early Fault Detection with Limited Data
}

\author{%
		Hamed Khorasgani\authorNumber{1}, Ahmed Farahat\authorNumber{2}, and Chetan Gupta\authorNumber{3}
}

\address{
	\affiliation{{1,2,3}}{Industrial AI Lab, Hitachi America, Ltd. R\&D, Santa Clara, CA, 95054, U.S.A.}{ 
		{\email{firstname.lastname@hal.hitachi.com}}
		} 
	
}

\maketitle

\phmLicenseFootnote{Khorasgani}

\begin{abstract}
Traditionally, fault detection and isolation community has used system dynamic equations to generate diagnosers and to analyze detectability and isolability of the dynamic systems. Model-based fault detection and isolation methods use system model to generate a set of residuals as the bases for fault detection and isolation. However, in many complex systems it is not feasible to develop highly accurate models for the systems and to keep the models updated during the system lifetime. Recently, data-driven solutions have received an immense attention in the industries systems for several practical reasons. First, these methods do not require the initial investment and expertise for developing accurate models. Moreover, it is possible to automatically update and retrain the diagnosers as the system or the environment change over time.  Finally, unlike the model-based methods it is straight forward to combine time series measurements such as  pressure and voltage with other sources of information such as system operating hours to achieve a higher accuracy. In this paper, we extend the traditional model-based fault detection and isolation concepts such as residuals, and detectable and isolable faults to the data-driven domain. We then propose an algorithm to automatically generate residuals from the normal operating data. We present the performance of our proposed approach  through a comparative case study.

\end{abstract}

\section{Introduction}
\begin{figure}[ht]
	\centering
	\includegraphics[scale=0.2]{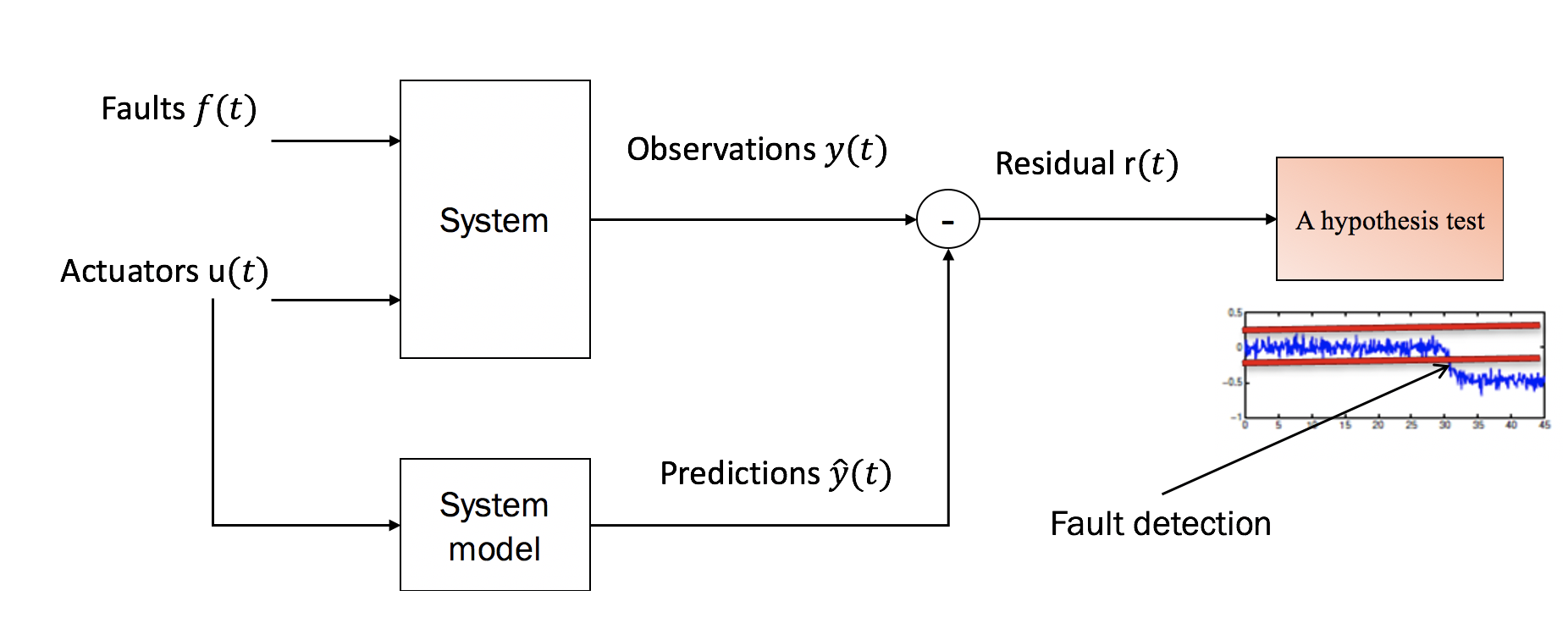}
	\caption{Model-based fault detection and isolation. At any given moment t, $u(t)$
represents the inputs from actuators and $y(t)$ represents the outputs
(measurements). The system faults are represented by  $f(t)$. The model represents the system' s digital twin and
the residuals, $r(t)$, are the differences between the system outputs and the model
outputs at time t.}
	\label{fig:fig_1}
\end{figure}

Deviations of system characteristics and parameters from standard conditions are referred to as faults in the
system \cite{isermann1997trends}. Faults can put the equipment operators at risk, disrupt the manufacturing processes and cost
industries millions of dollars. Fault detection determines the occurrence of a fault and the fault occurrence
time in the system. Subsequently, fault isolation determines the type and location of the detected fault \cite{isermann1997trends}.
Timely Fault Detection and Isolation (FDI) is critical for the system operators' safety and can help them to
prevent abnormal event progression and reduce downtime and productivity losses.
Traditionally, prognostics and health monitoring engineers use system models to monitor systems and
generate alarms when the system deviates from its normal operation. These methods simply compare the
system outputs to the model outputs to calculate the deviations from normal conditions. The differences
between the system outputs and the model outputs are called residuals. Residuals are the key components
in model-based FDI. To make an FDI method robust to noise and uncertainties, typically, a hypothesis test
such as Z-test \cite{biswas2003robust} is used to determine whether a residual deviation is statistically significant. Figure \ref{fig:fig_1} represents model-based fault detection and isolation.

In recent years, many research groups have expanded model-based fault detection and isolation to detect
and isolate more faults by using a broader definition for residuals and developing more sophisticated
approaches to derive the residuals \cite{mosterman1999diagnosis,bregon2013common,garcia1997deterministic}. A system model is a set of mathematical equations which
represent the system's normal behavior. A residual in the broader term is defined as an analytical
redundancy relationship (ARR) in the system equations. In this definition, a residual is not necessarily the
difference between the model output and the system output; instead any analytical relationship derived from the redundancies in the system is a residual. This expansion has several advantages. 1) It is been shown that
by using the analytical redundancy approach, we can detect and isolate faults which were not previously
detectable or isolable. 2) To design diagnosers using analytical redundancy approach, we do not need the
system complete model and a subset of equations may be adequate to detect and isolate the faults. 3) It is
possible to design more robust residuals using analytical redundancy approaches because they typically
include much less parameters, and measurements compared to the entire model.

Model-based methods are computationally efficient. Moreover, it is easy to understand and interpret the
diagnosis results of these approaches. However, for complex systems, developing reliable models can be
expensive, and it is often infeasible to derive a sufficiently accurate model for the system that generates
correct diagnosis results. When the system model is not available, data-driven solutions can be used for fault
detection and isolation. Instead of relying on the system models, data-driven solutions use system historical
data and machine learning techniques for FDI. There are two main data-driven fault detection and isolation
approaches: 1) classifier method and 2) system model methods \cite{salfner2010survey}.

The classifier approaches use both normal and fault data to train a classifier which classify each sample
point (or each time window which includes several successive samples points) as normal or fault modes.
When fault data is not available, system model methods are reliable alternatives. These methods use normal
data to learn system model and apply the trained model to compute expected value of the systems. When
the expected value does not match the current value, we can conclude that the system is not behaving as
expected and therefore, may be in a fault mode. Unlike classifier methods, system model methods do not
require fault data and therefore, are more practical.

In this paper, we propose a new data-driven ARR generation for fault detection and isolation as an
improved version of system model methods. Similar to the model-based approaches which use the system
physical equations for ARR generation, we use historical data during normal operation to extract
redundancies among sensor measurements. We call these redundancies data-driven ARR. We then use the
derived data-driven ARRs for fault detection and isolation. Like other data-driven methods, our solution
does not rely on the system model. Moreover, our solution can detect and isolate more faults than traditional
model learning methods, it can work when the system is not fully observable, and does not rely on vast
amount of historical fault data. This makes our approach practical in many real cases where there are data
limitations.

The rest of this paper is organized as follows. Section \ref{Prior Art} reviews the background and the previous work   in model-based and data-driven fault detection and isolations.
Section \ref{Problem Formulation} formally defines  the  data-driven ARR generation problem. Our proposed solution 
for generating  data-driven ARRs is presented in Section \ref{Methodology}. Section \ref{Case Study} demonstrates the application of our method through a case study.   The conclusions
are presented in  Section  \ref{Conclusions}.

\section{Prior Art and Background}
\label{Prior Art}

Residuals are fault indicators in model-based FDI methods \cite{isermann1997trends}. When a system is operating normally, the
residual values are expected to be close to zero. For example, consider a linear electric resistor. The
resistance can be measured as the ratio of electric voltage over electric current $R  = \frac{V}{I}$, where $R$ represents
the resistance, $V$ represents the electric voltage and $I$ represents the electric current. When both voltage and
current are measured, and we know the nominal value of the resistance, $R_n$, we have a redundancy in the
system. The difference between the measured resistance, $ \frac{V}{I}$  and nominal value of the resistance represents
a residual, $r= R_n - \frac{V}{I}$. During the normal operation $r$ is expected to be close to zero. However, if
because of for example a short-circuit fault the value of the resistance drops significantly, the residual is not
equal to zero anymore. Residuals are the most important part in any model-based fault detection and
isolation (FDI) solution. To detect a fault, $f$, model-based approaches require a residual sensitive to the fault
and, at the same time, invariant or at least robust to uncertainties and noise in the system. To isolate a fault
$f_i$ from another fault $f_j$ requires a residual sensitive to $f_i$ and at the same time insensitive to $f_j$ and other
uncertainties in the system \cite{frank1994frequency}.

Prognostics and health monitoring engineers use system models to extract residuals for fault detection and
isolation (FDI). A system's model is a set of mathematical equations which represent the system normal
behavior. A residual represents an analytical redundancy in system equations. Several approaches such as
graphical methods \cite{mosterman1999diagnosis,bregon2013common}, observer-based methods \cite{garcia1997deterministic}, and parity equations and ARRs \cite{gertler1998fault} have been
developed to extract residuals from system equations. ARR -based methods are among the most common
approaches for residual generation. These methods use two or more ways to determine the same variable, where at least one way uses model equations \cite{isermann1997trends}. A possible inconsistency between the two or more values
derived for the same variable is considered as a residual or fault indicator.

\begin{figure}[ht]
	\centering
	\includegraphics[scale=0.2]{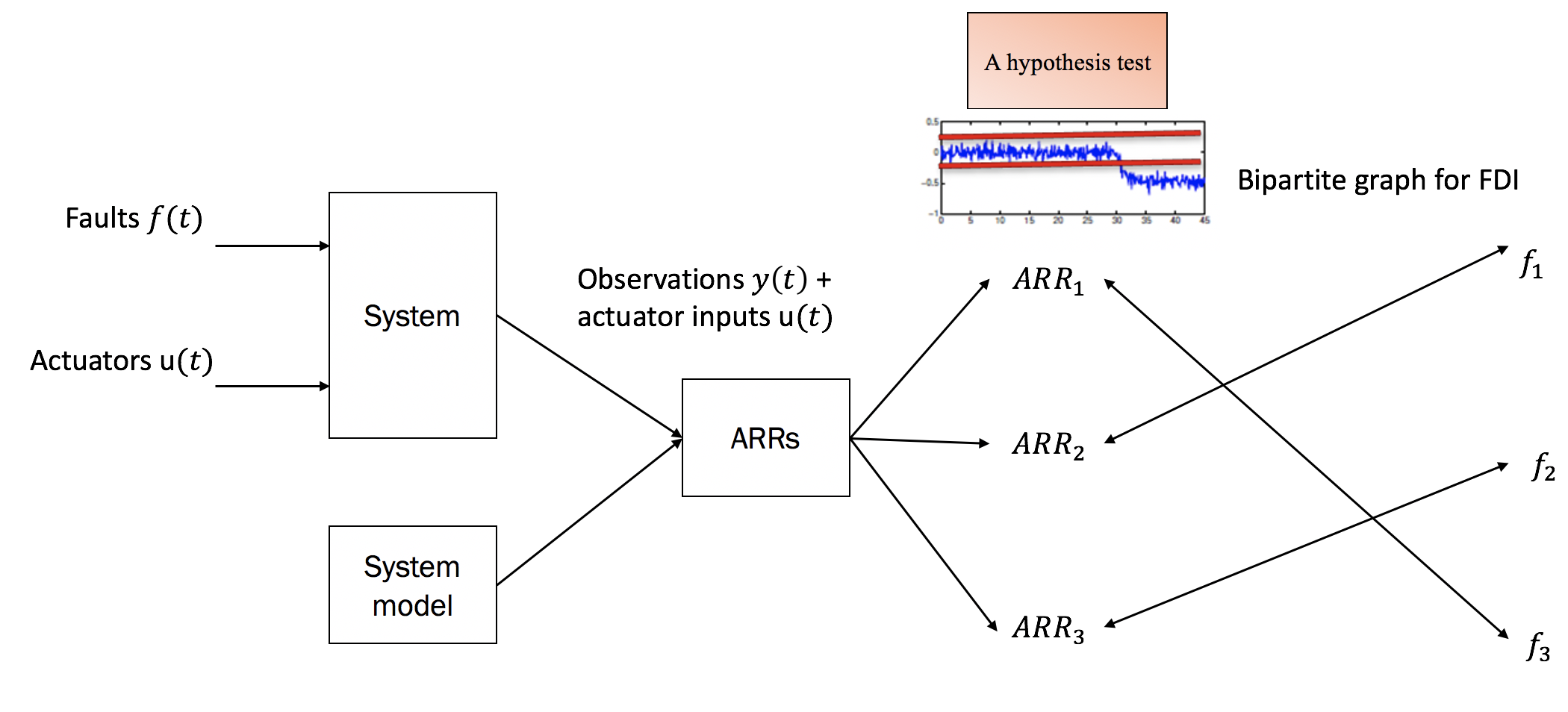}
	\caption{Model based FDI using ARRs.}
	\label{fig:fig_2}
\end{figure}

To make a diagnoser robust to noise and uncertainties, typically, a hypothesis test such as Z-test \cite{biswas2003robust} is used
to determine if a ARR deviation is statistically significant. In the last step, a fault isolation algorithm, uses
a decision logic to generate possible fault candidates based on the hypothesis tests outputs (alarms). Figure
\ref{fig:fig_2} represents a model-based fault detection and isolation scheme using ARRs.

In recent years data-driven diagnosis methods have been developed for fault detection and isolation \cite{venkatasubramanian2003review}. In
comparison, model-based methods have less computational costs. Moreover, it is easy to understand and
interpret model-based diagnosis results. However, for complex systems, developing reliable models can be
expensive, and it is often infeasible to derive a sufficiently accurate model for the system that generates
correct diagnosis results. Even when the model is available, it can become less reliable over time as the
system and the environment change gradually. In our example, the value of the resistance could change
because of the operating temperature. In this case, using the original model with the nominal value for FDI
leads to false alarms. Therefore, model based FDI is not practical in many real cases. When sufficiently
accurate models are not available, data-driven diagnosis methods bring promising alternatives to model-based
approaches \cite{venkatasubramanian2003review}. Instead of relying on the system models, data-driven solutions use system historical
data to learn models for FDI.

\begin{figure}[ht]
	\centering
	\includegraphics[scale=0.2]{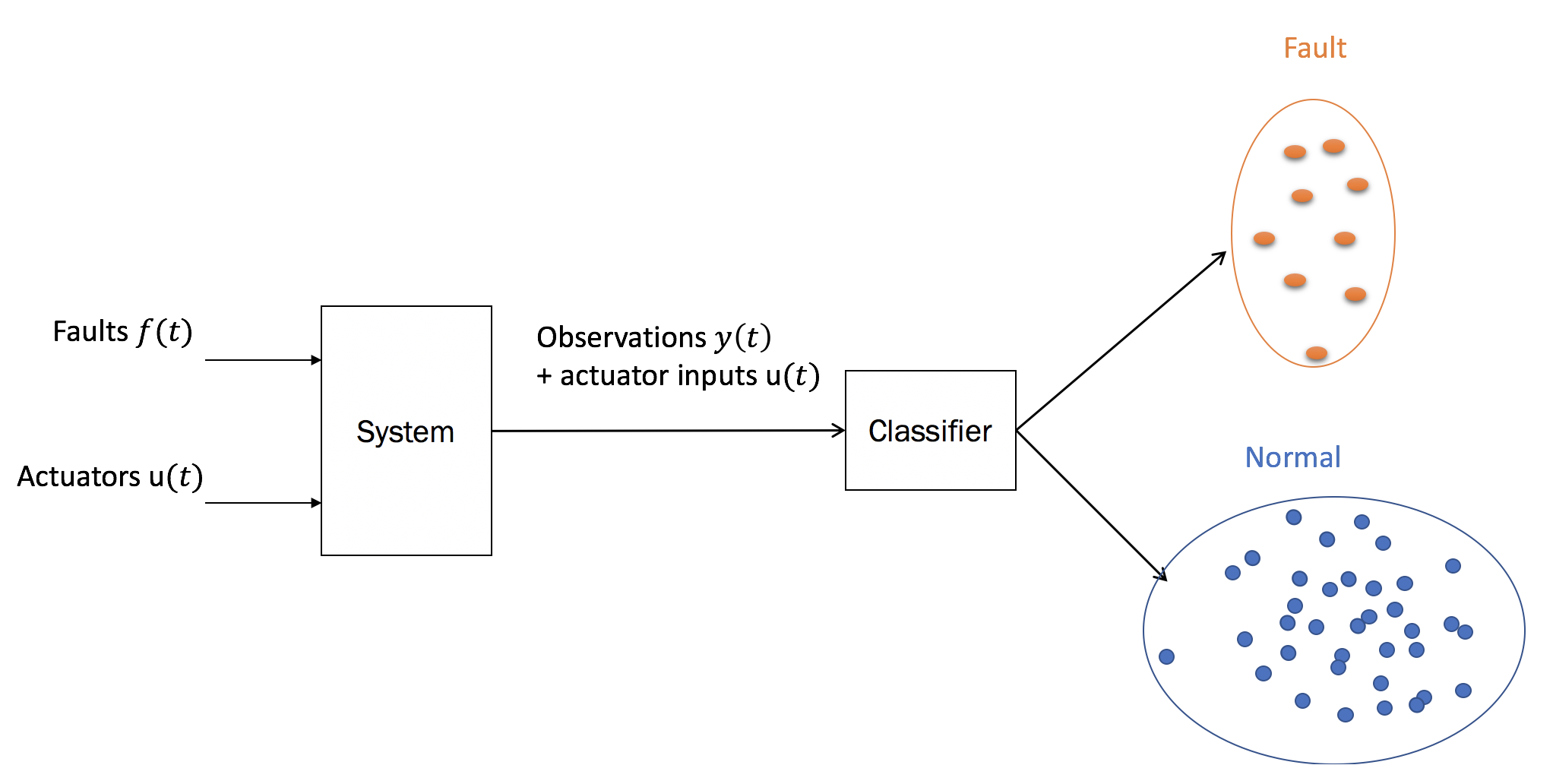}
	\caption{Data-driven classifier method fault detection.}
	\label{fig:fig_3}
\end{figure}

Data-driven FDI methods operate exclusively on measured data without detailed knowledge of the system.
Therefore, we do not need to have access to the system model. Moreover, as more data become available it
is possible to keep the model updated through retraining. There are two main data-driven FDI approaches:
1) classifier methods and 2) system model methods \cite{salfner2010survey}. Classifier method is shown in Figure \ref{fig:fig_3}. The classifier approaches address FDI in two steps; 1) feature selection and feature extraction and 2) fault classification.
The first step is designed to select a subset of relevant measurements or extract a set of new features from
the measurement data. Typically, this would represent a subset of measurements or extracted features that
are sensitive to the faults and, at the same time, invariant or at least robust to noise and disturbances in the
system. Among feature extraction methods, Principal Components Analysis (PCA) is the most widely used
\cite{venkatasubramanian2003review,wang2019remaining,wang2020health}. It generates a set of orthogonal bases in the directions where the data has the greatest variances. The
second step maps the features to the nominal operating mode or different fault modes. The classifier
approaches assume the training data is labeled by instances for the normal and fault classes. These methods
typically apply classifier methods such as neural networks or Bayesian networks to map the features to the
system fault modes. 

The classifier methods use both normal and fault data to train classifiers which classify each sample point
(or each time window which can include several successive samples points) as normal or fault modes. Data-driven
methods use normal and close to fault historical data to train their classifiers for fault detection. In
many domains, improvements in the production technology has led to more reliable systems. As the systems
become more reliable and less likely to fail, fewer historical fault data is available to train the classifiers. A
common solution is to generate fault data in the lab environment by using different methods such as
accelerated aging. These methods subject the system to high stresses to create different faults in a short
period of time in order to gather fault data for training. Clearly this approach can be very expensive
especially for complex systems.

\begin{figure}[ht]
	\centering
	\includegraphics[scale=0.35]{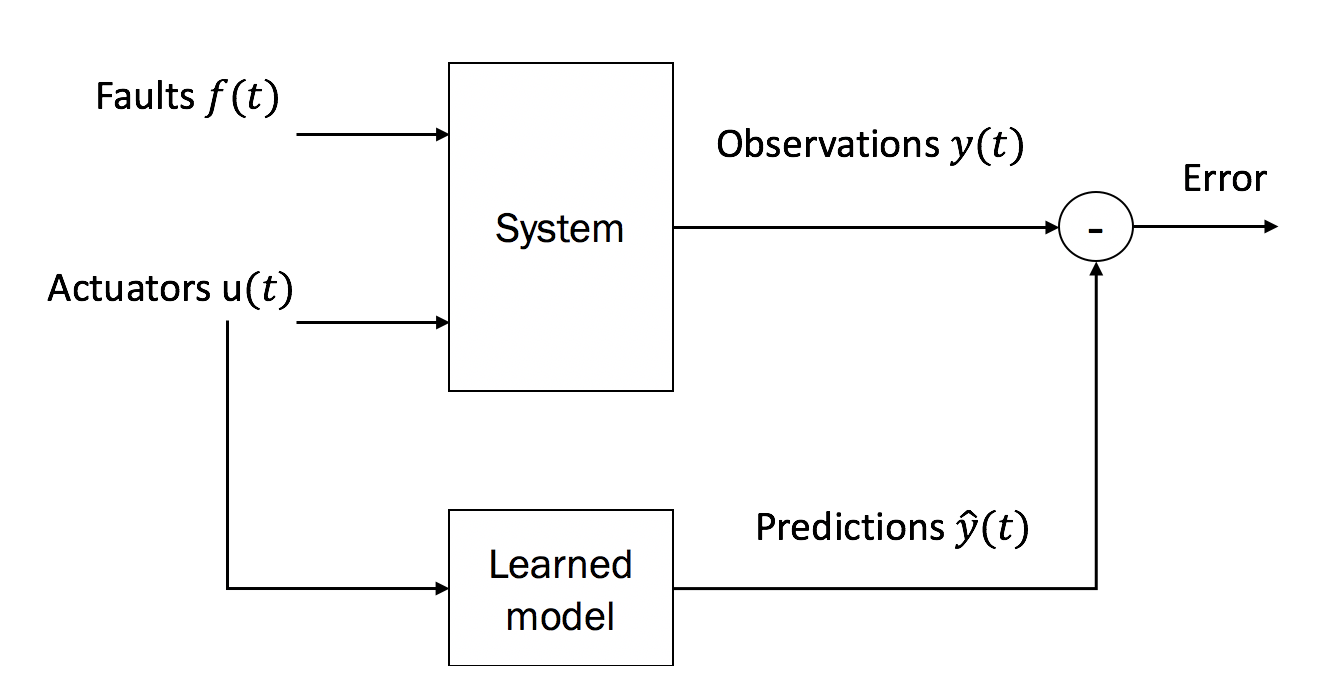}
	\caption{Data-driven system model fault detection methods.}
	\label{fig:fig_4}
\end{figure}

When fault data is not available, system model methods are reliable alternatives. These methods use normal
data to learn system model and apply the trained model to compute expected value of the systems. When
the expected value does not match the current value, we can conclude that the system is not behaving as
expected and therefore, may be in a fault mode. Unlike classifier methods, system model methods do not
require fault data and therefore, are more practical. A system model method is shown in Figure \ref{fig:fig_4}. 

In this paper, we argue that learning the system model from the data is not enough for accurate FDI and
we can achieve better results by learning the residuals from the data for the following reasons. 1) Model learning
approaches learn a model that maps system input to the system output. In many cases, the system input and
output are not obvious. Therefore, it is not trivial to learn the system overall model. In our residual
generation method, we do not define variables as input or output. Instead, we find a minimal set of variables
which can learn each target variable. 2) In many cases, we do not have access to all the measurements.
Therefore, we cannot learn the system model. However, we still may be able to learn residuals as they are
local sets of redundant variables. 3) In many cases it is challenging to isolate different faults in model
learning methods as they all affect the system output. Using the residuals, we can isolate different faults
much better. This is because of the local nature of the residuals. Several faults can affect different sensors.
By using minimal set of sensors in each residual we increase the likelihood of having a residual sensitive to
a fault and not sensitive to the other ones. This information can be used for fault isolation.

Instead of using the system physical model for residual generation, we use historical data during normal
operation to extract redundancies among sensor measurements. We call these redundancies data-driven
ARRs. We then use the data-driven ARRs for FDI. Like data-driven system model FDI methods, our
solution does not require system model and is not relied on a vast amount of historical fault data. However,
our solution can detect and isolate more faults than traditional model learning methods and can work when the system is not fully observable. These make our approach practical in many real cases where there are
data limitations.

\section{Problem Formulation}
\label{Problem Formulation}

In this section, we formally define the fault detection and isolation problem, data-driven ARRs, and
detectable and isolable faults using our data-driven method. These definitions help us formulate the data-driven
ARR generation problem in a systematic way.
\begin{definition}[Fault]
	\label{def:neighbors}
	A deviation of a system from standard condition is referred to as a fault.
\end{definition}

Early fault detection and isolation (FDI) is critical for the operators safety, and timely maintenance
recommendations which can save industries millions of dollars. We define FDI as follows.

\begin{definition} [FDI] Fault detection determines the occurrence of a fault and the fault occurrence time in the
system. In the next step, fault isolation pinpoints the type and location of the occurred fault in the system.
\end{definition}
Proposing a new method for data-driven ARRs is the main contribution of this paper. We define data-driven
ARRs as:
\begin{definition} [Data-driven ARR] A set of variables in the dataset, $V$, plus a target variable $v^t$ where $v^t \notin V$
 represent a data-driven ARR if there exists a machine learning model that can estimate $v^t$ using $V$ with
a given accuracy, $\epsilon$, for the validation data, $\hat{v}^t  = \text{model}(V)$, where $|| \hat{v}^t  - v^t|| \le \epsilon$ . The data-driven ARR for variable $v^t$ is
$r^t = \hat{v}^t - v^t$.
\end{definition}

In model-based analytical redundancy methods, we use minimal ARRs as the smallest subset of equations
which include redundancies for ARR generation. Minimal ARRs are more likely to be useful in fault
isolations (they are sensitive to fewer number of faults). Moreover, they tend to be less sensitive in model uncertainties as they include fewer parameters
and measurements. We define minimal data-driven ARRs as:

\begin{definition} [Minimal data-driven ARR] A data-driven ARR, $r = (v^t ,V)$ is a minimal ARR if no subset
of $V$ creates a data-driven ARR with the target variable $v^t$. 
\end{definition}
In model based FDI methods, we define a fault detectable when there exists a residual sensitive to the fault
and at the same time invariant or at least robust to uncertainties and noise in the system \cite{biswas2003robust}. Similarly, we define
a detectable fault using data-driven ARRs as:

\begin{definition} [Detectable fault] A fault $f$ is detectable if there exists a data-driven ARR,  $r = (v^t ,V)$, in the
data set where there is a statistically significant difference between $\hat{v}^t  = \text{model}(V)$ when fault $f$ occurs
and when it does not (normal operation).
\end{definition}

In model based FDI methods, we define a fault $f_i$ isolable from another fault $f_j$ when there exists a residual
sensitive to  $f_i$ and at the same time insensitive to $f_j$ and other uncertainties in the system \cite{biswas2003robust}. Similarly, we
define isolable faults using data-driven ARRs as:

\begin{definition} [Isolable faults] Fault $f_i$ is isolable from fault $f_j$ if there exists a data-driven ARR, $r = (v^t ,V)$,
in the data set where there is significant statistical difference between $\hat{v}^t  = \text{model}(V)$ when fault $f_i$ occurs
and when fault $f_j$ occurs.
\end{definition}

Using these definitions, we develop a method to generate data-driven ARRs and design a data-driven
diagnoser for FDI in the next section. 

\section{Methodology}
\label{Methodology}
In this section, we propose two algorithms for data-driven ARR generation, and two methods to address
delay in dynamic systems. In the next section, we present a simple case study  to clarify our solution.

\subsection{Data-driven Residual Generation Algorithm}
In the previous section, we defined a data-driven ARR as a set of variables in the dataset, $V$, plus a target
variable $v^t$  where we can use $V$ to estimate $v^t$. To find these set of variables for each target variable, we
propose two method 1) exhaustive search, 2) forward feature selection. The exhaustive search algorithm
can find several ARRs for each target variable. This method can be helpful when there are several faults in
the system and more ARRs can help isolating faults from each other. However, the exhaustive search is
computationally expensive. The feature selection algorithm is an efficient alternative solution. The feature
selection method finds at most one ARR per target variable. Even though this may lead to missing some
ARRs, we believe feature selection algorithm is sufficient for most applications.

\textit{Exhaustive search:} In the exhaustive search method, for each target variable  $v^t$ we find all the minimal set of variables that can
estimate  $v^t$. Each of these variable groups plus  $v^t$ is a minimal data-driven ARR. We use a tree search
algorithm to find these residuals. 
\begin{itemize}
\item For each target variable $v^t$, we start with all the variables, and see if they can estimate $v^t$ accurately
(there exists a machine learning model that can estimate $v^t$ using the variables with the required
accuracy). If they cannot, it means there is no residual in the dataset for $v^t$ as even all the variables
cannot estimate $v^t$. In this case, we move to the next variable.
\item If they can, all the variables plus $v^t$ create an ARR. However, we cannot be sure that this ARR is
minimal. For each variable in $V$, we remove the variable and check if the subset still can estimate
$v^t$, if no subset of $V$  can estimate $v^t$, it means $V$ plus $v^t$ is a minimal ARR. We save this ARR and
move to the next variable.
\item Otherwise, the ARR is not minimal. In this case, we repeat this procedure for every subset of
variables that can estimate $v^t$, till we find all the minimal ARR. In more details, for each variable
$v \in V$ we remove $v$ and check if $V - v$ plus $v^t$ is an ARR. If not, we know there is no ARR in $V - v$  for
$v^t$, if yes, we check if it is also minimal. If the data-driven ARR is minimal, we save $V - v$  plus $v^t$ as
a new ARR. Otherwise, keep removing variables till we reach a minimal ARR.
\item After finding all the minimal ARRs for $v^t$, we move to the next variable.
\end{itemize}

Note that to check if a set of variables can estimate the target variable, we use the selected set of variables
to learn a model $\hat{v}^t = \text{model}(V)$ and compute estimation score for the learned model. The model can be
linear regression, neural network, support vector regression or any other regression model based on the
application.

\textit{Forward feature selection:} The exhaustive search finds all the minimal data-driven ARRs in the dataset and therefore, it is guaranteed
to achieve maximum detectability and isolability. However, this algorithm is computationally expensive
and may not be practical for systems with large number of measurements. To address this problem, we
propose a forward feature selection algorithm as an alternative solution. We use forward feature selection
to select the minimum number of variables which can estimate the value of each variable in the dataset.
\begin{itemize}
\item For each target variable, $v^t$, in the set of variables we go through the dataset and measure the
estimation score for each variable in the dataset. We add the variable with the highest estimation
score as the first variable in the residual list, $R$.
\item We then go through all the remained variables and add the variable which achieves the highest improvement in  estimation
score to the residual list, $R$. We stop when we reach to the required accuracy score, $\text{error}  \le
\epsilon$, where $\hat{v}^t = \text{model}(R)$, and $\text{error} = ||v_t - \hat{v}^t ||$. If we do not reach the required accuracy
score, we conclude there is no residual for $v^t$.
\item At the end, we go through all the variables selected for each residual and remove the ones which
have no significant contribution to the overall set of variables.
\end{itemize}

The data-driven ARR for variable $v^t$  is $r = v^t - \text{model}(R)$. Just because this relationship holds, this does
not mean that $r$ is useful in fault detection and isolation. In the next subsection, we find a subset of
generated residuals that are useful for fault detection and isolation. Toward this end, we use statistical
analysis such as Z-test \cite{biswas2003robust}to see if the generated residuals are statistically different with and without faults
and therefore, can be used for FDI.

\subsection{Selecting Useful ARRs}
Not all the analytical redundancies (data-driven ARRs) are useful for fault detection and isolation. For each
fault mode in the system, we go through all the ARRs and select the best set of ARRs to detect the fault.
Different metrics can be used to measure the performance of ARRs in fault detection and isolation. For the
sake of demonstration, we use the following simple approach:
For each residual we use Z-test to quantify the difference between the residual in faulty and normal dataset. We select the residuals that their hypothesis  tests show statistically significant difference between
normal and faulty data. The selected residuals can also be
used to isolate faults.  

\begin{algorithm}
\small
	\def \MMM {\ensuremath{\mathcal{M}}\xspace}
	\def \DDD {\ensuremath{\mathcal{D}}\xspace}
	\def \UUU {\ensuremath{\mathcal{U}}\xspace}
	\newcommand{\Set}[1]{\ensuremath{\{#1\}}}
	\caption{Data-driven Residual Generation}
	\label{alg:Res}
	\begin{algorithmic}[1]
		\STATE {Let M be the set of measurements in the system. }
		\STATE {Let $\Delta$ be the set of possible delays in the system. }
		\STATE {Initialize the  set of residuals, $R =  \emptyset$. }
		\FOR{$m_r  \in  M$ }
		\STATE {Initialize the set of loads for $m_r$, $L_{r}= \{ \}$. }
		\STATE {${r}_{\text{score}}= 0$. }
		\FOR{iteration = 1:$N_l$}
		\STATE {Initialize $m\text{candidate} =  \emptyset$. }
		\FOR{$m_l  \in M$ }
		\IF{$m_l\ne m_r $}
		\FOR{$\Delta t  \in \Delta$ }
		\STATE {Use normal data to train  $m_r = \text{model}(L_{r} + \{ m_l(t-\Delta t ) \} )$.  }
		\IF{$\text{model}_{\text{score}} > {r}_{\text{score}} + \epsilon$}
		\STATE {$m_{\text{candidate}} =  \{ m_l(t-\Delta t ) \}$ }
		\STATE {${r}_{\text{score}}=\text{model}_{\text{score}}$}
		\ENDIF
		\ENDFOR
		\ENDIF
		\ENDFOR
		\STATE {$L_{r}=L_{r} +  m_{\text{candidate}} $ }
		\ENDFOR
		\STATE {Use normal data to train  $m_r = \text{model}(L_{r}  )$.  }
		\STATE {Use normal data to calculate   $r_{n}= m_r - \text{model}(L_{r}  )$ .  }
		\STATE {Use failure  data to calculate   $r_{f}= m_r - \text{model}(L_{r}  )$ .  }
		\IF{there is a statistically significant difference between $r_{n}$ and $r_{f}$ }
		\STATE { $R = R + \{(m_r, L_{r} ) \}$. }
		\ENDIF
		\ENDFOR

	\end{algorithmic}
\end{algorithm}    

\subsection{Delays in the System}
\begin{figure}[ht]
	\centering
	\includegraphics[scale=0.35]{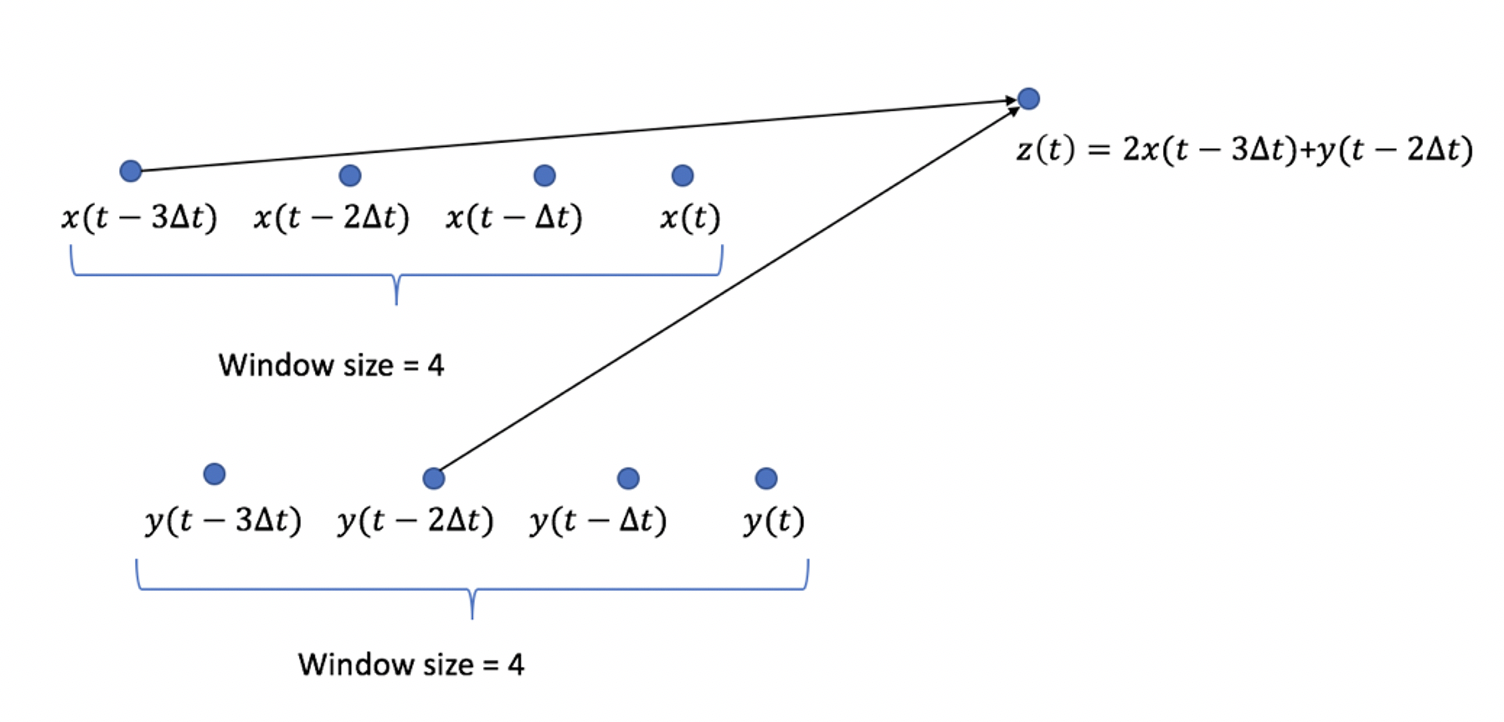}
	\caption{variables with delays.}
	\label{fig:fig_5}
\end{figure}

In many cases the variables are correlated, and they represent redundancies in the system. However, because
of delays in dynamic systems for example the time it takes for heat to transfer through materials, we cannot
use them to estimate target variable if we only use the current sample values. We propose two solutions to
this problem 1) using window size 2) using sequential models such as recurrent neural networks (RNNs). By using a window of last n
samples for each variable, our model can capture delays in the system and generate data-driven ARRs for
systems with delays. Figure \ref{fig:fig_5} shows a simple example where variable z is a function of variables x, and y.

However, because of the delay in the system the effect of variable x on z appears with $3 \Delta t$ and the effect of variable y appears with $2 \Delta t$ delay. As it is shown in the figure, we can capture this residual by using a window of size 4 for each variable when estimating the target variable. The window size is a meta parameter which users can define based on their knowledge of the system. Algorithm \ref{alg:Res}  shows the overall solution.

It is also possible to use sequential models such as recurrent neural networks (RNNs) and long-short term memory (LSTM) to address delay. RNNs include a memory loop in their structure and therefore, they can use information from previous samples in estimating the target variable. RNNs have been used in many applications such as speech recognition, and language modeling in recent years. Long Short-Term Memory networks (LSTMs) are a subclass of RNNs, which are capable of learning long-term dependencies as well \cite{hochreiter1997long}. When long-term delays exist in the system, and we expect long-term dependencies among the variables we can use LSTM to learn the residuals. However,  training LSTM networks could be computationally expensive, and the users should consider computational limitations when selecting this model.

\section{Case Study}
\label{Case Study}
In this case study, we show our proposed data-driven residual generation approach can generate a set of residuals for a 4 tank system \cite{khorasgani2019structural} using historical data. The 4 tank system includes the following measurements:
\begin{figure}[h]
	\centering
	\includegraphics[scale=0.29]{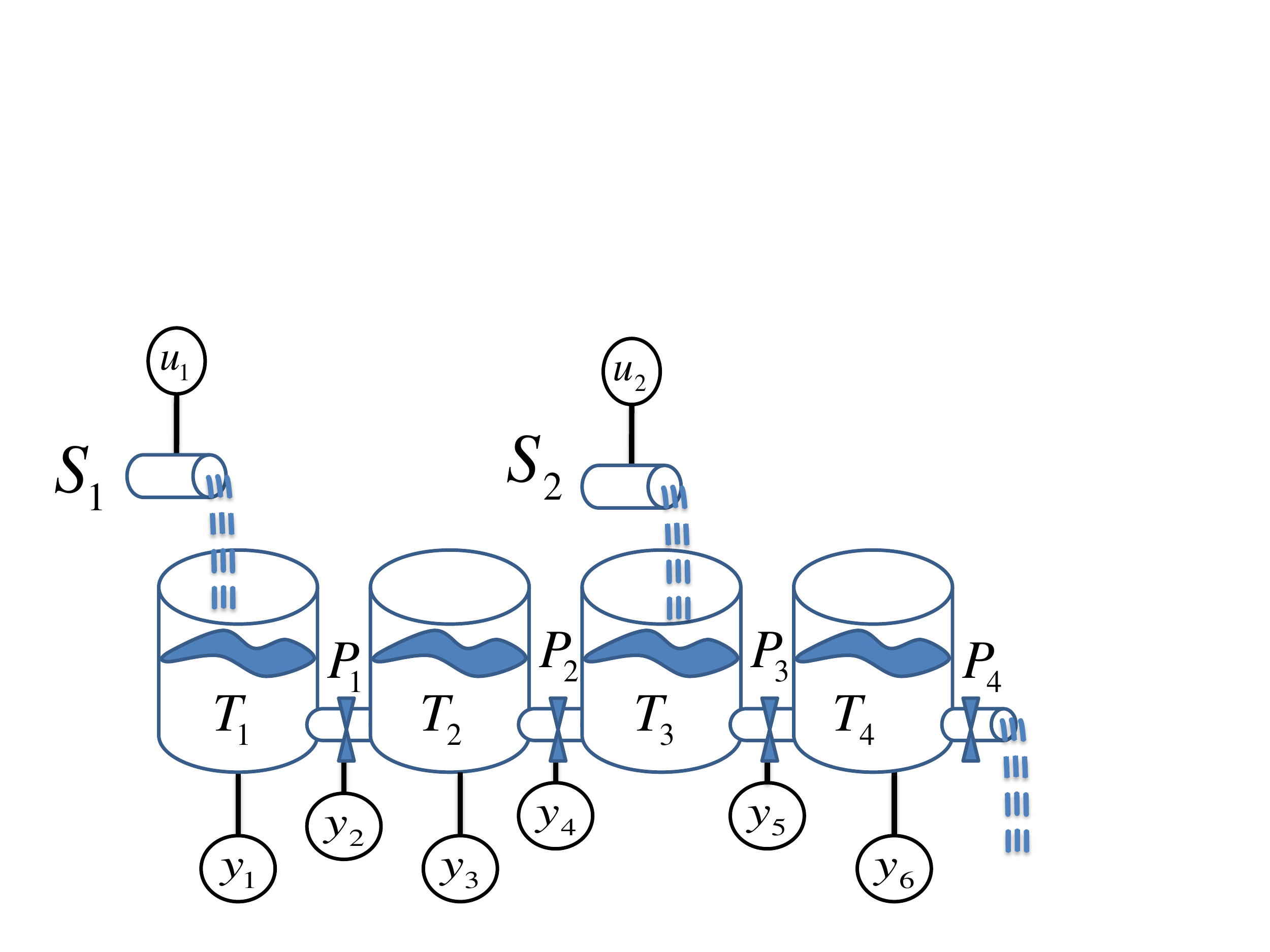}
	\caption{Case study: four tank system.}
	\label{fig:Case Study}
\end{figure} 
\begin{equation}
\begin{aligned}[c]
\begin{split}
 q_{in1} &=  u_1  \\ 
{p}_1 &=  y_1 \\
 {q}_1 &=   y_2\\
 {p}_2&=  y_3 \\
\end{split}
\end{aligned}
\qquad \qquad
\begin{aligned}[c]
\begin{split}
 {q}_2 &=   y_4 \\
 q_{in2} &=  u_2  \\
q_{3} &=  y_5  \\
 p_{4} &=  y_6  \\
\end{split}
\end{aligned}
\label{eq:measurements}   
\end{equation}
where $q_{in1}$ is the flow rate of the inflow to tank 1,  $p_1$ is the pressure in tank 1, ${q}_1$ is the flow rate of the outflow from tank 1 to tank 2,  $p_2$ is the pressure in tank 2, ${q}_2$ is the flow rate of the outflow from tank 2 to tank 3, $q_{in2}$ is the flow rate of the inflow to tank 3, $q_{3}$ is the flow rate of the outflow from tank 3 to tank 4, and $p_4$ is the pressure in tank 4. Note that some of the flow rates and pressure variables in the system such as pressure in tank 3 and flow rate out of tank 4 are not measured. 

\begin{figure}[ht]
	\centering
	\includegraphics[scale=0.35]{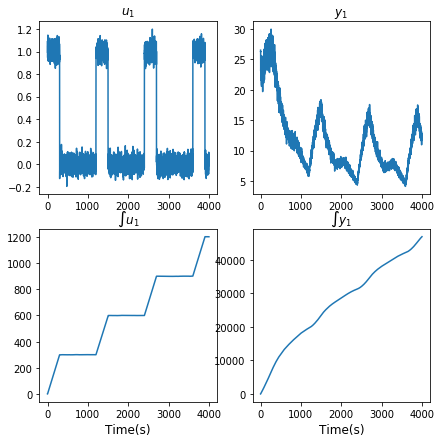}
	\caption{$u_1$ and $y_1$ and their integrals. There is  $5\%$ gaussian noise in each measurement. We use $SciPy$ library in python to compute the integrals.}
	\label{fig:fig_int}
\end{figure}

For dynamic systems such as the four tank system, we consider the integral of each variable as a new variable to generate residuals. This  increases the number of  residuals and can improve the diagnosability. We can also use the derivative of variables. However, the derivative is highly sensitive to noise. To calculate the  integrals, we need the initial variables.  Typically, this is the main motivation to choose derivative casualty in model-based methods \cite{frisk2012diagnosability}. An inaccurate  initial  value  adds a bias to the integral variable. In the data-driven approach,  a bias can be easily learned by machine learning models and does not cause any challenge. Figure \ref{fig:fig_int} shows $u_1$ and $y_1$ and their integrals.

\begin{table*}[ht]
	\caption{Generated residuals using model-based vs data-driven methods} 
	\centering 
	\begin{tabular}{||c|| c |c |c } 
		\hline\hline 
		Target Variable & Selected Variables & Residual\\ 
		\hline 
		$u_1$ & $\times$  & $\times$\\
		\hline
		$y_1$ & $y_2$, $y_3$ & $r_{y1}=y_1-\text{model}(y_2,y_3)$ \\
		
		\hline
		$u_2$ &$\times$ & $\times$ \\
		
		\hline
		$y_2$ & $\times$ & $\times$\\
		
		\hline
		$y_3$ & $y_1$, $y_2$ & $r_{y_3}=y_3-\text{model}_{y_3}(y_1,y_2)$ \\
		
		\hline
		$y_4$ & $y_3$, $y_5$ , $y_6$& $r_{y_4}=y_4-\text{model}_{y_4}(y_3,y_5,y_6)$ \\
		
		\hline
		$y_5$ &$\times$ & $\times$  \\
		\hline 
		$y_6$ &$\times$ & $\times$  \\
		\hline %
		$\int u_1$ & $y_1, \int y_2$  & $r_{\int u_1}=\int u_1-\text{model}(y_1, \int y_2)$\\
		\hline
		$\int y_1$ & $\times$  &  $\times$  \\
		
		\hline
		$ \int u_2$ &$\times$ & $\times$ \\
		
		\hline
		$\int y_2$ & $\times$ & $\times$\\
		
		\hline
		$\int y_3$ &  $\times$ &  $\times$ \\
		
		\hline
		$\int y_4$ &  $\times$ &  $\times$  \\
		
		\hline
		$\int y_5$ &$\times$ & $\times$  \\
		\hline 
		$\int y_6$ &$\times$ & $\times$  \\
		\hline %
	\end{tabular}
	\label{table:results} 
\end{table*}
Using our proposed residual generation algorithm based on forward feature selection, we can find at most one data-driven ARR for each variable. Table \ref{table:results} shows the generated residual for each target variable. $\times$ means our forward feature selection approach has failed to find any residual for the given target variable. Note that forward feature selection method is a greedy algorithm and may miss some  of the residuals. We applied linear regression as our machine learning model and used  the coefficient of determination, $R^2>99\%$ as a measure to determine if the machine learning model can estimate the target value. 

Some of the derived residuals are trivial, for example we know that the flow rate between two tanks is a function of the difference between their pressures, therefore, it is not surprising to see the model extracts the relationship between $p_1 = y_1$, $q_1=y_2$ and $p_2=y_3$ to generate residuals $r_{y_1}$ and $r_{y_3}$. However, in some other cases our algorithm was able to derive more complicated relationships. For example, consider residual $r_{y_4}$. We know that ${q}_2 =   y_4$ is a function of pressure in tank 2, $p_2$ and pressure in tank 3, $p_3$. Even though $p_3$ is not among the measurements, our algorithm has been able to use pressure in the next tank   $p_{4} =  y_6$ to estimate $y_4$. As an other example, $ r_{\int u_1}$ could capture the dynamic relationship between the inflow to tank one, its pressure and the outflow from tank 1 to tank 2.

\begin{figure}[ht]
	\centering
	\includegraphics[scale=0.35]{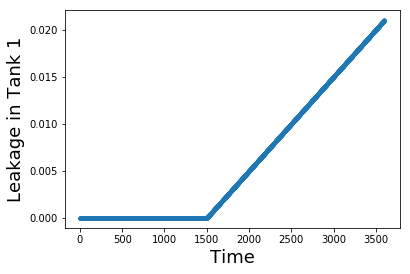}
	\caption{ An incipient fault in tank 1.}
	\label{leakage}
\end{figure}
\begin{figure}[ht]
	\centering
	\includegraphics[scale=0.35]{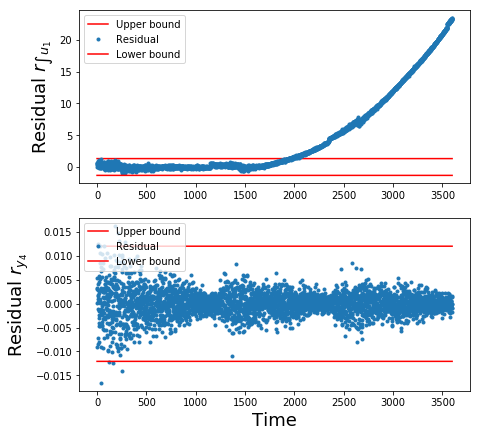}
	\caption{ Data-driven residuals in the presence of the leakage fault.}
	\label{residuals}
\end{figure}

 
As we discussed in this paper, not all the residuals are useful in detecting an specific fault. For example, consider a leakage in tank 1 modeled as an  incipient fault as it is shown in Figure \ref{leakage}. Algorithm \ref{alg:Res} finds $r_{\int u_1}$ as the useful residual to detect this fault.
Figure \ref{residuals} shows that residual $r_{\int u_1}$ is sensitive to this fault but residual $r_{y1}$ is not. The upper and lower bounds for each residual are derived using residual value during normal operation, $r_n$:  $\text{upper bound}=\hat{r}_n+3 \sigma{r_n} $, $\text{lower bound}=\hat{r}_n- 3 \sigma{r_n} $.

\begin{figure}[ht]
	\centering
	\includegraphics[scale=0.35]{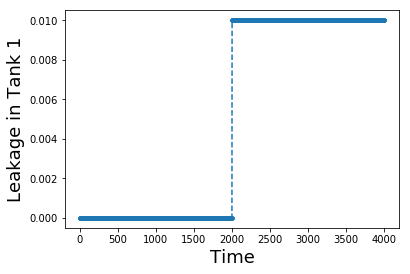}
	\caption{An abrupt fault in tank 1.}
	\label{abrupt}
\end{figure}

The main advantage of data-driven residual based fault detection and isolation  compared to traditional data driven classifier based methods is that the data-driven residual generation requires no fault data for residual generation and limited fault data for evaluating the residuals. In addition, the data-driven residuals magnify the effect of fault in sensor data, and therefore, they can reduce the detection time, and improve overall accuracy. For example, consider the following scenario where we have access to data with an abrupt leakage in Tank 1 during the training.  Figure \ref{abrupt} shows the abrupt fault. 

\begin{figure}[ht]
	\centering
	\includegraphics[scale=0.35]{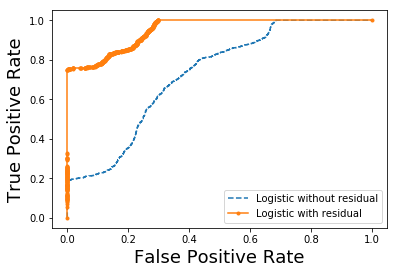}
	\caption{Receiver Operating Characteristic (ROC) curve for leakage detection with and without the selected residual, $r_{\int u_1}$.}
	\label{final_resualts}
\end{figure}

 To show the the advantage of using data-driven residuals compared to relying purely on sensor data, we designed the following experiment. We use the data with abrupt failure to train a logistic regression model and we use this model to detect the incipient fault shown in Figure \ref{leakage}. In the first model we only use sensor variables form the 4-tank system. In the second model we use the sensor data +  selected residual, $r_{\int u_1}$. Figure \ref{final_resualts} shows the data-driven residual significantly improves the model performance. 

\section{Conclusions}
\label{Conclusions}
In this paper, we proposed a new data-driven ARR generation for fault detection and isolation. Our approach uses historical data during normal operation to extract redundancies among sensor measurements. We call these redundancies data-driven ARRs. We then use the derived data-driven ARRs for fault detection and isolation. Our solution has the following advantages:
1) It does not rely on the system model.
2) It can detect and isolate more faults than traditional data-driven methods. 3) It can work when the system is not fully observable.
4) It does not rely on vast amount of historical fault data.
These advantages make our approach practical in many real cases where there are data limitations.

\bibliographystyle{apacite}
\bibliography{ijphm}

\end{document}